\documentclass[aps,prd,preprint,groupedaddress,nofootinbib]{revtex4}
\usepackage{amsmath,amssymb,graphics,graphicx,epsfig,subfigure}

\newcommand{\bsym}{\boldsymbol}
\newcommand{\eps}{\epsilon}
\newcommand{\newc}{\newcommand}
\newc{\beqa}{\begin{eqnarray}}
\newc{\eeqa}{\end{eqnarray}}
\newc{\beq}{\begin{equation}}
\newc{\eeq}{\end{equation}}
\newc{\ra}{\rightarrow}
\newc{\tri}{\triangle}

\begin{document}

\title{Shadow Higgs from a scale-invariant hidden $U(1)_s$ model}
\author{We-Fu Chang}
\email{wfchang@phys.nthu.edu.tw} \affiliation{Department of
Physics, National Tsing-Hua University, Hsinchu 300, Taiwan}
\author{John N. Ng}
\email{misery@triumf.ca}
\author{Jackson M. S. Wu}
\email{jwu@triumf.ca}
\affiliation{Theory group, TRIUMF, 4004 Wesbrook Mall, Vancouver, B.C., Canada}

\date{\today}

\begin{abstract}
We study a scale invariant $SU(2)\times U(1)_Y \times U(1)_s$ model which has
only dimensionless couplings. The shadow $U(1)_s$ is hidden, and it interacts 
with the Standard Model (SM) solely through mixing in the scalar sector and 
kinetic mixing of the $U(1)$ gauge bosons. The gauge symmetries are broken 
radiatively by the Coleman-Weinberg mechanism. Lifting of the flat direction 
results in a light shadow Higgs or ``scalon'', and a heavier scalar which we 
identify as the SM Higgs boson. The phenomenology of this model is discussed. 
It is possible that shadow Higgs boson can be discovered in precision $t$-quark
studies at the LHC. The conditions that it be a dark matter candidate is also
discussed.
\end{abstract}

\pacs{}

\maketitle

\section{Introduction}
There has been much recent interest in the idea of Standard Model (SM)
having a hidden sector wherein the matter content are SM gauge singlets, but
transform non-trivially according to the hidden sector gauge
groups~\cite{SW05,CFW05,SZ06a,PW06,MW06,CDU06,OCRMW06,BTGR06,MN06,EQ07}. Hidden
sectors arise in many top-down models, including those inspired by the brane
world scenario and string theory. Most discussion of them posit an association
with a very high mass scale, and their couplings to the visible SM sector are
often through nonrenormalizable or loop effects. This need not be the case
however, and it has been noticed that through renormalizable interactions, the
hidden sector can be probed at energies soon to be available at the Large
Hadron Collider (LHC).

We consider here a simple case where the hidden sector contains a single
complex scalar gauged under an additional $U(1)$ to the hypercharge of the SM.
Such a  $U(1)$ factor is ubiquitous in gauge theories as it forms a part of a
more complicated gauge group. Thus, we expect the physics we explore in this
paper would be generic across a variety of models.

There are two gauge-invariant (and renormalizable) ways the $U(1)$ gauged
hidden sector can communicate with the SM fields. One is through kinetic
mixing between the field strengths of the SM $U(1)_Y$ and our hidden sector
``shadow'' $U(1)_s$. In older constructions where the extra gauge sector
couples directly to the SM fermions, this leads to the well-known $Z'$
physics~\cite{ExtraZ}. In the hidden sector context, there is no direct
coupling, and the phenomenological impact of the gauge mixing between $U(1)_Y$
and $U(1)_s$ has been studied in~\cite{EL99,ADH02,KW06,CNW06}. The other way
is through mixing between the SM Higgs with the hidden sector scalar, the
``shadow Higgs'' $\phi_s$. In this paper we examine the phenomenology of this
Higgs mixing in a complete model.

Motivated by the anticipated start up of LHC, models studying the
modifications of the SM Higgs signal due to an extended Higgs sector uncharged
under the SM gauge group have been proposed (see e.g.~\cite{BHR06,MW06}). The
additional scalars are often constructed to be heavier (if not very much so)
than the SM Higgs to avoid the current bounds from electroweak precision tests
(EWPTs). But with a hidden sector construction, this need not be so. Indeed,
if the hidden sector scalars are very light ($\lesssim 100\,\mathcal{O}$(MeV)),
they can be candidates for dark matter~\cite{BF04} (under suitable
assumptions).

With this in mind, we focus in this paper on a special case of the
renormalizable model given in~\cite{CNW06} where it is classically conformal
invariant, and the symmetry breaking is induced radiatively via
Coleman-Weinberg (CW) mechanism~\cite{CW73}~\footnote{Similar ideas has
previously been applied in the context of grand unified theory resulting in a
different phenomenology~\cite{Hempf96}. CW mechanism in a hidden sector
context have recently also been applied in the dynamical generation of the
neutrino mass~\cite{MN06} and the electroweak phase transition~\cite{EQ07}.}.
Besides its elegance, the occurrence of a small mass scale through CW
mechanism is a natural consequence of the conformal symmetry breaking. This
feature precluded the implementation of CW mechanism in a SM context because
the prediction of the Higgs mass there ($\lesssim 10$~GeV) is far lower than
the current lower bound of $114.4$~GeV at $95\%$~CL from direct searched at
LEP2~\cite{LEP2}. But in terms of our hidden $U(1)_s$ model with its one extra
scalar, the same feature becomes key in ensuring in addition to a SM-like
Higgs boson, a light shadow Higgs, which is not only viable under the current
EWPT constraints, it also generates a new signal in the top decays testable at
the LHC. In analyzing our model, we apply the methods of Gildener and S.
Weinberg (GW)~\cite{GW76} which allows perturbation theory to be used in a CW
context with multiple scalars.

The organization of our paper is as follows. In the next section, we review
the work of GW to set up the framework. In Sec.~\ref{Sec:Model} we apply the
methods of GW to our model and calculate the mass of the scalar bosons. In
Sec.~\ref{Sec:HiggsPhen}, we discuss the phenomenology of the scalar sector
of our model. We summarize in Sec.~\ref{Sec:Summ}.

\section{Review of GW results}\label{Sec:RevGW}
In this section, we review the main idea of the GW analysis, and we record
useful formulae from that work to set up the framework from which we apply to
our model. We will follow Ref.~\cite{GW76} closely below.

The work of GW is the earliest comprehensive study of the effective potential
that extended the analysis of CW to massless field theories with multiple
scalar fields. They considered a renormalizable gauge theory with an arbitrary
multiplet of real scalar fields $\Phi_i$. The tree level potential is given by
\begin{equation}\label{Eq:potn}
V_0(\Phi) = \frac{1}{24}f_{ijkl}\Phi_i\Phi_j\Phi_k\Phi_l \,.
\end{equation}
Typically, the nonzero components of $f_{ijkl}$ are of order $e^2$, where
$e \ll 1$ stands for the generic gauge couplings in the theory.

To ensure that perturbation theory will stay valid throughout the analysis,
the prescription of GW is to choose a value $\Lambda_W$ of the renormalization
scale $\Lambda$, at which $V_0(\Phi)$ has a nontrivial minimum on some
ray $\Phi_i = N_i\,\phi$, where $\bsym{N}$ is a unit vector in the field space
and $\phi$ is the radial distance from the origin of the field space. This
prescription is implemented by adjusting $\Lambda$ so that
\begin{equation}\label{Eq:fcond}
\min_{N_i N_i = 1}V_0(\bsym{N}) =
\min_{N_i N_i = 1}f_{ijkl}N_i N_j N_k N_l = 0 \,.
\end{equation}
Note that this imposes only a \textit{single} constraint on the $f_{ijkl}$.
One cannot choose a renormalization scale such that all $f_{ijkl}$ vanish,
just a single combination.

Suppose the minimum~\eqref{Eq:fcond} is attained for some specific unit vector
$N_i = n_i$. Then one necessary condition is that of the stationary point
\begin{equation}\label{Eq:statcond}
\frac{\partial V_0(\bsym{N})}{\partial N_i}\biggr|_{\bsym{n}} = 0
\Longleftrightarrow f_{ijkl}\,n_j n_k n_l = 0 \,.
\end{equation}
For $V_0(\bsym{N})$ to attain a minimum at $\bsym{N} = \bsym{n}$ further
requires that for all vectors $\bsym{u}$
\begin{equation}\label{Eq:Pmtx}
P_{ij}u_i u_j \geq 0 \,, \qquad
P_{ij} \equiv \frac{\partial^2 V_0(\bsym{N})}{\partial N_i\partial N_j}
\biggr|_{\bsym{n}} = \frac{1}{2}f_{ijkl}\,n_k n_l \,,
\end{equation}
i.e. the eigenvalues of $P$ are either positive or zero .

Turning on the higher-order corrections $\delta V$ in the potential will give
rise to a small curvature in the radial direction, which picks out a definite
value $v$ of $\phi$ at the minimum, as well as causing a small shift in the
direction of the ray $\Phi_i$ at this minimum. The stationary point condition
at the new, perturbed minimum $\bsym{n}v + \delta\Phi$ is
\begin{equation}
0 = \frac{\partial}{\partial\Phi_i}
\big(V_0(\Phi) + \delta V(\Phi)\big)\biggr|_{\bsym{n}v + \delta\Phi} \,,
\end{equation}
or to first order in small quantities
\begin{equation}\label{Eq:pMinDir}
0 =  P_{ij}\,\delta\Phi_j\,v^2 +
\frac{\partial\delta V(\Phi)}{\partial\Phi_i}\biggr|_{\bsym{n}v} \,.
\end{equation}
This uniquely determines $\delta\Phi$ except for possible terms in directions
along eigenvectors of $P$ with eigenvalue zero, which includes $\bsym{n}$ by
construction
\begin{equation}\label{Eq:scalonC}
P_{ij}\,n_j = \frac{1}{2}
\frac{\partial V_0(\bsym{N})}{\partial N_i}\biggr|_{\bsym{n}} = 0 \,,
\end{equation}
and the Goldstone modes $\Theta_\alpha\bsym{n}$ corresponding to the continuous
symmetries $\Theta_\alpha$. There is no reason, in general, to expect any
other eigenvectors of $P$ with zero eigenvalues, and is assumed so.

Instead of using~\eqref{Eq:pMinDir} to determine $\delta\Phi$,
contracting~\eqref{Eq:pMinDir} with $n_i$ and using~\eqref{Eq:scalonC} leads
to a basic equation that determines the value of $v$
\begin{equation}\label{Eq:MinVEV}
0 = \frac{\partial}{\partial\Phi_i}\delta V(\Phi)\biggr|_{\bsym{n}v} =
\frac{\partial}{\partial\phi}\delta V(\bsym{n}\phi)\biggr|_v \,.
\end{equation}
Calculating $\delta V$ to one-loop, the potential along the ray
$\Phi = \bsym{n}\phi$ can be written in the form
\begin{equation}\label{Eq:V1LAB}
\delta V(\bsym{n}\phi) = A\,\phi^4 + B\,\phi^4\log\frac{\phi^2}{\Lambda_W^2}
\,,
\end{equation}
where $A$ and $B$ are dimensionless constants
\begin{align}
A &= \frac{1}{64\pi^2 v^4}\left\{
3\mathrm{Tr}\!\left[M_V^4\log\frac{M_V^2}{v^2}\right]
+\mathrm{Tr}\!\left[M_S^4\log\frac{M_S^2}{v^2}\right]
-4\mathrm{Tr}\!\left[M_F^4\log\frac{M_F^2}{v^2}\right]\right\} \\
B &= \frac{1}{64\pi^2 v^4}\left(
3\mathrm{Tr}\,M_V^4 + \mathrm{Tr}\,M_S^4 - 4\mathrm{Tr}\,M_F^4\right) \,.
\end{align}
The trace is over all internal degrees of freedom, and $M_{V,S,F}$ are the
zeroth-order vector, scalar, and spinor mass matrices respectively, for
a scalar field vacuum expectation value $\bsym{n}v$.

From~\eqref{Eq:V1LAB}, the stationary point condition~\eqref{Eq:MinVEV}
implies
\begin{equation}\label{Eq:vVEV}
\log\frac{v^2}{\Lambda_W^2} = -\frac{1}{2}-\frac{A}{B} \,.
\end{equation}
Because of the choice of the renormalization scale~\eqref{Eq:fcond}, both
$A$ and $B$ are of order $e^4$, so the logarithm is of order unity, and
perturbation theory should be valid. Note that this implies $\Lambda_W$ and
$v$ are of the same order.\footnote{See~\cite{GW76} for more details.}

The squared masses of the scalar bosons are given by the eigenvalues of the
second derivative matrix of the effective potential
\begin{equation}\label{Eq:Msqto1L}
(M^2)_{ij} = (M_0^2 + \delta M^2)_{ij} =
\frac{\partial^2}{\partial\Phi_i\partial\Phi_j}
\big[V_0(\Phi) + \delta V(\Phi)\big]\biggr|_{\bsym{n}v + \delta\Phi} \,,
\end{equation}
where
\begin{equation}\label{Eq:MsqLO}
(M_0^2)_{ij} = \frac{\partial^2 V_0(\Phi)}{\partial\Phi_i\partial\Phi_j}
\biggr|_{\bsym{n}v} = P_{ij}\,v^2 \,,
\end{equation}
is the zeroth order scalar mass-squared matrix, and
\begin{equation}\label{Eq:MsqNLO}
(\delta M^2)_{ij} =
\frac{\partial^2\delta V(\Phi)}{\partial\Phi_i\partial\Phi_j}
\biggr|_{\bsym{n}v} + f_{ijkl}\,n_k\,\delta\Phi_l\,v \,,
\end{equation}
to first order in small quantities, with $\delta\Phi$ determined
from~\eqref{Eq:pMinDir}.

From the discussion above, $M_0^2$ has a set of positive-definite
eigenvalues of order $e^2 v^2$ corresponding to Higgs bosons, plus
a set of zero eigenvalues with eigenvectors
$\Theta_\alpha\bsym{n}$ corresponding to Goldstone bosons, plus
one zero eigenvalue with eigenvector $\bsym{n}$, the ``scalon''.
Provided that $\delta V$ has the same symmetries as $V_0$ and is a
small perturbation, the Higgs boson mass would remain
positive-definite , and the Goldstone bosons would remain
massless.

The scalon is a pseudo-Goldstone boson arising from the spontaneous symmetry
breaking of the conformal symmetry. Its mass can be straightforwardly
calculated from first-order perturbation theory
\begin{equation}
m_s^2 = n_i\,n_j\,(\delta M^2)_{ij} =  n_i\,n_j\,
\frac{\partial^2\delta V(\Phi)}{\partial\Phi_i\partial\Phi_j}
\biggr|_{\bsym{n}v} =
\frac{\partial^2}{\partial\phi^2}\delta V(\bsym{n}\phi)\biggr|_{\bsym{v}} \,.
\end{equation}
From~\eqref{Eq:V1LAB} and~\eqref{Eq:vVEV}, this gives
\begin{equation}\label{Eq:mssq}
m_s^2 = 8B\,v^2 \,.
\end{equation}

\section{The conformal shadow model and its breaking}\label{Sec:Model}
The complete Lagrangian of our model takes the form~\cite{CNW06}
\begin{equation}
\mathcal{L} = \mathcal{L}_{SM}
-\frac{1}{4}X^{\mu\nu}X_{\mu\nu}-\frac{\epsilon}{2}B^{\mu\nu}X_{\mu\nu}
+\left|\left(\partial_{\mu}-\frac{1}{2}g_s X_{\mu}\right)\phi_s\right|^2
-V_0(\Phi,\phi_s) \,.
\end{equation}
We consider here an initially scale-invariant theory in which the tree level
scalar potential is given by
\begin{equation}\label{Eq:V0}
V_0(\Phi,\phi_s) =
\lambda(\Phi^\dag\Phi)^2 + \lambda_s(\phi_s^*\phi_s)^2
+2\kappa\left(\Phi^\dag\Phi\right)\left(\phi_s^*\phi_s\right) \,.
\end{equation}
We assume that the quartic coupling constants $\lambda$, $\lambda_s$, and
$\kappa$ are all of order at least $g_s^2$, where $g_s \ll 1 $ is the gauge
coupling constant of the shadow $U(1)_s$.

In unitary gauge, the scalar fields on some ray $\varphi_i = \rho\,N_i$, where
$\bsym{N}$ is a unit vector in the field space $\{\Phi\otimes\phi_s\}$, can be
parameterized as
\begin{equation}\label{Eq:GWCoord}
\Phi = \frac{\rho}{\sqrt{2}}
\begin{pmatrix}
0 \\
N_1
\end{pmatrix} \,, \qquad
\phi_s = \frac{\rho}{\sqrt{2}}N_2 \,.
\end{equation}
In terms of these coordinates, the tree level potential has the form
\begin{equation}
V_0(\bsym{\varphi}) = V_0(\rho,\bsym{N}) =
\frac{\rho^4}{4}(\lambda N_1^4 + \lambda_s N_2^4 + 2\kappa N_1^2 N_2^2) \,.
\end{equation}
We assume that $\lambda$ and $\lambda_s$ are positive so that the potential
is bounded below.

The GW condition~\eqref{Eq:fcond} and~\eqref{Eq:statcond} that $V_0$ attains
a minimum value of zero on a unit sphere for some unit vector
$\bsym{N} = \bsym{n}$ implies that
\begin{equation}
\frac{\partial V_0}{\partial N_i}\biggr|_{\bsym{n}} = 0 \,, \qquad
V_0\bigr|_{\bsym{n}} = 0 \,.
\end{equation}
The solution of these equations is given by
\begin{equation}\label{Eq:MinDir}
n_1^2 = \frac{\sqrt{\lambda_s}}{\sqrt{\lambda}+\sqrt{\lambda_s}} \,, \qquad
n_2^2 = \frac{\sqrt{\lambda}}{\sqrt{\lambda}+\sqrt{\lambda_s}} \,, \qquad
\kappa = -\sqrt{\lambda\lambda_s} \,.
\end{equation}
The first two relations specify the direction of the unperturbed minimum of
the zeroth-order potential $V_0$; the last relation is a consistency condition
that $V_0$ vanishes along this direction.

Along the ray $\varphi_i = n_i\rho$, the one-loop effective potential is
given by
\begin{equation}\label{Eq:shV1LAB}
V_{1L}(\bsym{n}\rho) = A\,\rho^4 + B\,\rho^4\log\frac{\rho^2}{\Lambda_W^2} \,,
\end{equation}
where
\begin{align}
\label{Eq:shA}
A &=
\frac{1}{64\pi^2 v^4}\biggl\{
6m_W^4\log\frac{m_W^2}{v^2} + 3m_{Z_1}^4\log\frac{m_{Z_1}^2}{v^2}
3m_{Z_2}^4\log\frac{m_{Z_2}^2}{v^2} \notag \\
&\qquad\qquad\qquad +
m_{H,\,0}^4\log\frac{m_{H,\,0}^2}{v^2} - 12m_t^4\log\frac{m_t^2}{v^2}
\biggr\} \,, \\
\label{Eq:shB}
B &=
\frac{1}{64\pi^2 v^4}\left(
6m_W^4 + 3m_{Z_1}^4 + 3m_{Z_2}^4 + m_{H,\,0}^4 - 12m_t^4\right) \,.
\end{align}
Note that we have included only the $t$-quark contribution since it overwhelms
all other fermionic contributions.

The mass of the vector bosons at tree level are given by
\begin{align}
\label{Eq:mWsq}
m_W^2 &= \frac{1}{4}g_W^2 n_1^2 v^2 = \frac{1}{4}g_W^2 v_r^2 \,, \\
\label{Eq:mZ12sq}
m_{Z_{1,\,2}}^2 &= \frac{v^2}{8}\Biggl\{
n_1^2\Bigl[g_W^2 + g_Y^2(1 + s_\eps^2)\Bigr] + c_\eps^2 n_2^2\,g_s^2 \notag \\
&\qquad\qquad \mp
\sqrt{4\,c_\eps^2 s_\eps^2\,n_1^2 n_2^2\,g_Y^2 g_s^2 +
\Bigl[n_1^2\,(g_W^2 + g_Y^2(1 + s_\eps^2)) - c_\eps^2 n_2^2\,g_s^2\Bigr]^2}
\Biggr\} \notag \\
&= \frac{v_r^2}{8}\Biggl\{g_s(r,\eps)^2 + g_W^2 + g_Y^2(1 + s_\eps^2)
\notag \\
&\qquad\qquad \mp \sqrt{4\,s_\eps^2\,g_Y^2 g_s(r,\eps)^2 +
\Bigl[g_W^2 + g_Y^2(1 + s_\eps^2) - g_s(r,\eps)^2\Bigr]^2}\Biggr\} \,,
\end{align}
where~\footnote{The signs here correct the typographical error in the
definition of the same quantities in~\cite{CNW06}.}
\begin{equation}
s_\eps = \frac{\eps}{\sqrt{1 + \eps^2}} \,, \qquad
c_\eps = \frac{1}{\sqrt{1 + \eps^2}} \,,
\end{equation}
and we have defined
\begin{equation}\label{Eq:rdef}
r \equiv \frac{\sqrt{\lambda}}{\sqrt{\lambda_s}} \,, \qquad
v_r \equiv n_1 v = \frac{v}{\sqrt{1 + r}} \,, \qquad
g_s(r,\eps) \equiv c_\eps\sqrt{r}g_s \,.
\end{equation}
Note that we work in the mass-diagonal basis where the gauge kinetic terms are
in canonical form, and these are the gauge bosons in that basis.

The mass of the scalar boson at tree level is given by
\begin{equation}\label{Eq:mH0sq}
m_{H,\,0}^2 = 2\sqrt{\lambda\lambda_s}\,v^2 \,.
\end{equation}
There is only one heavy Higgs boson in our model that has a tree level mass
which is given $m_{H,\,0}$. The only other massive scalar boson is the scalon,
but it has no tree level mass. The scalon gets its mass purely from radiative
processes through CW mechanism, and is light. From~\eqref{Eq:shB},
\eqref{Eq:mWsq}, \eqref{Eq:mZ12sq}, and~\eqref{Eq:mH0sq}, the scalon mass as
defined in~\eqref{Eq:mssq}, is given by
\begin{align}\label{Eq:shmssq}
m_s^2 &= 8B\,v^2 \notag \\
&= \frac{3v_r^2}{64\pi^2(1 + r)}\left[
\frac{3g_W^4}{2} + g_Y^2 g_W^2(1 + s_\eps^2) +
\frac{g_Y^4}{2}(1 + s_\eps^2)^2 +
s_\eps^2 g_Y^2 g_s(r,\eps)^2 + \frac{g_s(r,\eps)^4}{2}\right] \notag \\
&\qquad + \frac{v_r^2}{2\pi^2}(1 + r)\,\kappa^2
-\frac{3m_t^4}{2\pi^2 v_r^2(1 + r)} \,.
\end{align}

After spontaneous breaking of conformal symmetry by the CW mechanism, we can
write the scalar fields as
\begin{equation}
\Phi = \frac{1}{\sqrt{2}}
\begin{pmatrix}
0 \\
n_1 v + h
\end{pmatrix} \,, \qquad
\phi_s = \frac{1}{\sqrt{2}}(n_2 v + s) \,,
\end{equation}
where $h$ and $s$ are the excitations about the minimum along directions
$n_1$ and $n_2$ respectively. From~\eqref{Eq:V0}, the tree level potential
$V_0$ then takes the form
\begin{align}
V_0(\Phi,\phi_s) &=
\frac{\lambda}{4}\,h^4 + \lambda\,n_1 v\,h^3 + \kappa\left(
n_2 v\,h^2 s + \frac{1}{2}\,h^2 s^2 + n_1 v\,h\,s^2\right) +
\lambda_s\,n_2 v\,s^3 + \frac{\lambda_s}{4}\,s^4 \notag \\
&\quad +
\frac{v^2}{2}(3n_1^2\,\lambda + n_2^2\,\kappa)\,h^2 +
2\kappa\,n_1 n_2 v^2\,h\,s +
\frac{v^2}{2}(n_1^2\,\kappa + 3n_2^2\,\lambda_s)\,s^2 \,,
\end{align}
with the linear terms vanish by~\eqref{Eq:MinDir}.

The physical mass-diagonal basis is defined by
\begin{equation}\label{Eq:physH}
\begin{pmatrix}
h \\
s
\end{pmatrix}
= U
\begin{pmatrix}
H_1 \\
H_2
\end{pmatrix}
=
\begin{pmatrix}
n_1\,H_1 - n_2\,H_2 \\
n_2\,H_1 + n_1\,H_2
\end{pmatrix} \,,
\end{equation}
where $U$ is an orthogonal matrix given by
\begin{equation}\label{Eq:Umtx}
U =
\begin{pmatrix}
n_1 & -n_2 \\
n_2 &  n_1
\end{pmatrix}
=
\begin{pmatrix}
\frac{1}{\sqrt{1 + r}} & -\frac{\sqrt{r}}{\sqrt{1 + r}} \\
\frac{\sqrt{r}}{\sqrt{1 + r}} & \frac{1}{\sqrt{1 + r}}
\end{pmatrix} \,.
\end{equation}
Note that the matrix $U$ is exactly the matrix which diagonalize the zeroth
order scalar mass matrix $M_0^2$ as defined in~\eqref{Eq:MsqLO} (or
equivalently, the matrix $P$ defined in~\eqref{Eq:Pmtx})
i.e.\footnote{Because of our choice of the unitary gauge~\eqref{Eq:GWCoord}
rotated away all gauge degrees of freedom, $M_0^2$ will contain no zero
eigenvalues corresponding to Goldstone bosons.}
\begin{align}
M_0^2 &= P\,v^2 =
\begin{pmatrix}
3n_1^2\lambda + n_2^2\,\kappa & 2n_1 n_2\,\kappa  \\
2n_1 n_2\,\kappa & n_1^2\kappa + 3n_2^2\lambda_s
\end{pmatrix}v^2 \notag \\
&= U^{-1}
\begin{pmatrix}
m_{H_1}^2 & 0 \\
0 & m_{H_2}^2
\end{pmatrix}
U
= U^{-1}
\begin{pmatrix}
0 & 0 \\
0 & 2\sqrt{\lambda\lambda_s}
\end{pmatrix}
U\,v^2 \,.
\end{align}
We see thus that $H_1$ corresponds to the scalon state, and $H_2$ corresponds
to the heavy Higgs boson state.

Going to the physical basis, we get with the help of~\eqref{Eq:MinDir}
and~\eqref{Eq:rdef}
\begin{align}\label{Eq:FeynR}
V_0(\Phi,\phi_s) &=
\frac{m_{H,\,0}^2}{2}H_2^2 -
\sqrt{\frac{\lambda}{2}}\left(1-\frac{1}{r}\right)m_{H,\,0} H_2^3 +
\frac{\lambda}{4}\left(1-\frac{1}{r}\right)^2 H_2^4 \notag \\
&\quad -\kappa\,H_1^2 H_2^2 - 2\kappa\sqrt{1 + r}\,v_r\,H_1 H_2^2 -
\sqrt{\lambda|\kappa|}\left(1-\frac{1}{r}\right)H_1 H_2^3 \,.
\end{align}
Note that from~\eqref{Eq:MinDir}, $\kappa < 0$ since $\lambda,\,\lambda_s > 0$.
The Feynman rules in the scalar sector of our model can be readily read off
from~\eqref{Eq:FeynR}.

Notice in~\eqref{Eq:FeynR} quartic terms contain no more than two scalon
($H_1$) fields, and in cubic terms no more than one. This is a general feature
of the GW framework that follows from the stationary point
condition~\eqref{Eq:statcond}: Recall the general form of the tree level
potential~\eqref{Eq:potn}. After symmetry breaking, the scalar field takes the
form
\begin{equation}
\Phi_i = n_i v + \varphi_i = n_i v + (U\cdot\bsym{H})_i \,, \qquad
(U\cdot\bsym{H})_i = \sum_i\bsym{\zeta_i}H_i \,,
\end{equation}
where $\varphi_i$ are the excitations about the minimum in the $i$-th
direction, and $\bsym{\zeta}_i$ are the eigenvectors of the mass matrix
$M_0^2$. By construction, the flat direction $\bsym{n}$ is always an
eigenvector of $M_0^2$ (see discussion above). Thus, since the scalon is by
definition the state associated with $\bsym{n}$, the statement follows.

\section{Constraints and phenomenology}\label{Sec:HiggsPhen}
The parameters relevant for the scalar sector in our model are the gauge
kinetic mixing angle $s_\eps$, the gauge coupling $g_s$, the quartic couplings
$\lambda$, $\lambda_s$, $\kappa$, and the scalar field vacuum expectation
value $v$. Because of relations~\eqref{Eq:MinDir} and~\eqref{Eq:rdef}, we will
trade in $\lambda_s$ and $v$, and use the parameter set
$\{s_\eps,\,g_s,\,\lambda,\,\kappa,\,r,\,v_r\}$ for convenience of analysis
below.

From the mass of the $W$ boson~\eqref{Eq:mWsq}, and the relation between
$m_W$ and the Fermi coupling constant
\begin{equation}
\frac{G_F}{\sqrt{2}} = \frac{g_W^2}{8m_W^2} \,,
\end{equation}
$v_r$ can be determined, and it is given by
\begin{equation}
v_r = 2^{-1/4}\,G_F^{-1/2} = 246.221\,\mathrm{GeV} \,.
\end{equation}

Since our interest here is in exploring the parameter space of the scalar
sector of our model, given that $s_\eps \lesssim 10^{-2}$ (see
Ref.~\cite{CNW06}), we will neglect higher order corrections in $\eps$, and
treat $s_\eps$ as zero in the analysis below~\footnote{For the physical,
parity-even processes we consider below, the leading corrections start at
$\mathcal{O}(\eps^2)$. We set $s_\eps = 0$ here purely for the purpose of
simplifying the analysis; the kinetic mixing parameter $\eps$ should never be
thought of as being identically zero, as then it implies a complete decoupling
of the shadow Z from the visible sector that would upset the cosmological
bounds. We leave the more complete but also more complicated analysis that
kept $s_\eps \neq 0$ at leading order to future works~\cite{Future}.}.

\begin{figure}[htbp]
\centering
\includegraphics[width=4in]{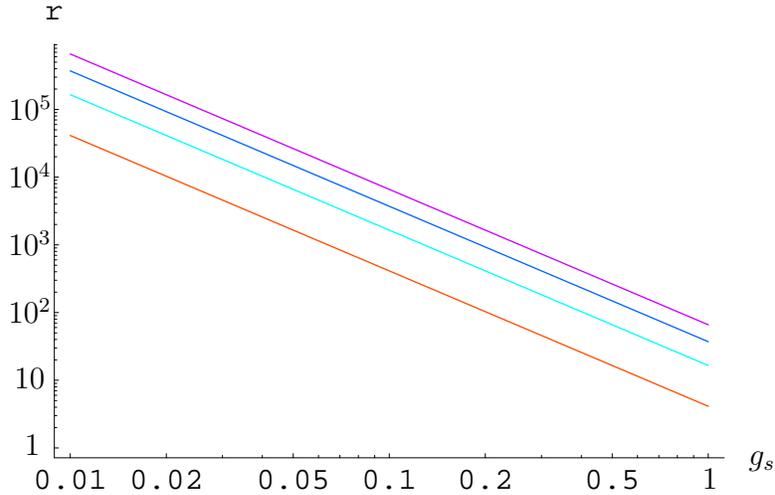}
\caption{\label{Fig:rgsplot}
The parameter $r$ as a function of $g_s$ for $m_{Z_s}$ fixed at
$250$~(bottom line), $500$, $750$, and $1000$~(top line)~GeV.}
\end{figure}

\begin{figure}[htbp]
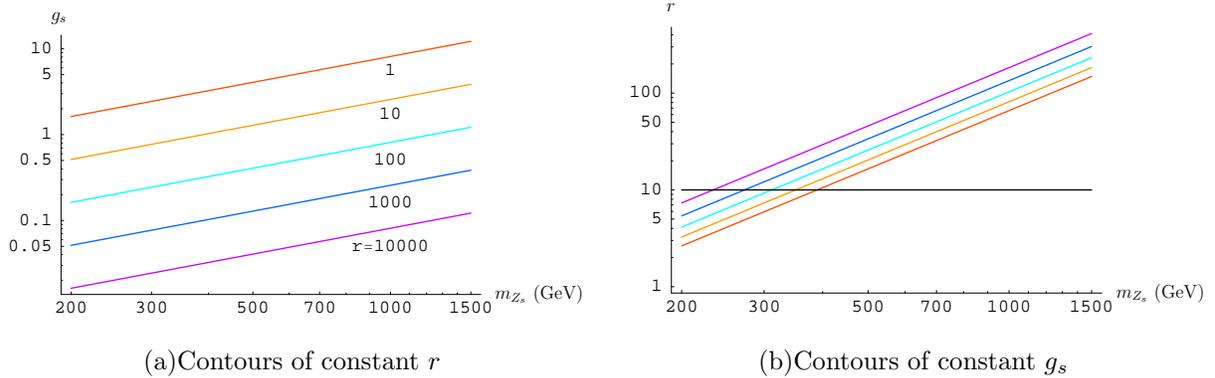

\centering
\subfigure[Contours of constant $r$]{
\label{subfig:rgsmZsC}
\includegraphics[width=3in]{rgsmZs.epsi}}
\hspace{0.1in}
\subfigure[Contours of constant $g_s$]{
\label{subfig:rmZsgs1}
\includegraphics[width=3in]{rmZs_gs1.epsi}}
\caption{\label{Fig:rmZsplot}
Relationships between $r$, $g_s$, and $m_{Z_s}$. In~\subref{subfig:rmZsgs1},
the contour lines are for fixed value of $g_s = 1$ (bottom line), $0.9$, 
$0.8$, $0.7$, and $0.6$ (top line).}
\end{figure}

Setting $s_\eps = 0$, we get from~\eqref{Eq:mZ12sq}
\begin{equation}
m_{Z_1}^2 = \frac{v_r^2}{4}(g_W^2 + g_Y^2) = m_Z^2 \,, \qquad
m_{Z_2}^2 = \frac{v_r^2}{4}\,r\,g_s^2 \,,
\end{equation}
i.e., $Z_1$ is automatically the SM $Z$, while $Z_2$ is the shadow
$Z$~\footnote{The convention we adopt is that $Z_2$ shall always denote the
heavier state, viz. the shadow $Z$. With $s_\eps = 0$, this corresponds to
taking the negative sign for the square roots in~\eqref{Eq:mZ12sq}.}.
With $v_r$ fixed, we can write $r$ as a function of $g_s$ and $m_{Z_2}$
($= m_{Z_s}$)
\begin{equation}\label{Eq:rse0}
r = \frac{4m_{Z_2}^2}{v_r^2}\frac{1}{g_s^2}
=\frac{m_{Z_s}^2}{m_W^2}\frac{g_W^2}{g_s^2} \,.
\end{equation}
In Fig~\ref{Fig:rgsplot} and ~\ref{Fig:rmZsplot}, we show the functional
interdependencies implied by~\eqref{Eq:rse0}.

As is apparent from Eq.~\eqref{Eq:rse0}, Fig.~\ref{Fig:rgsplot} and
Fig.~\ref{Fig:rmZsplot}, for a given value of $g_s$, the heavier the shadow
Z, the larger $r$ needs to be. For example, if $g_s \sim g_W \sim 0.65$, to
have $m_{Z_s} > 500$~GeV would require $r > 10^2$, and $r$ would be even
larger if $g_s$ is smaller. In Fig.~\ref{subfig:rmZsgs1}, each contour line
forms a lower bound on the values of $r$ for a given value of $g_s$. We see
from it that if $r$ is of order $10$, for $m_{Z_s} > 1$~TeV, $g_s$ would have
to exceed its perturbative limit, which we conservatively take to be
$g_s = 1$. While for $500$~GeV~$< m_{Z_s} < 1$~TeV, $g_s$ would have to be
close to the unity.

We now turn our attention to the Higgs sector. In contrast to the multiscalar
construction with CW mechanism in a GUT context, we emphasize and reiterate
here that it is the heavy scalar boson, $H_2$, that will take the role of the
SM Higgs boson (with SM Higgs mass) in our model. Being a pseudo-Goldstone 
boson of the spontaneously broken conformal symmetry, the scalon, $H_1$, or 
the ``shadow Higgs''  is naturally light, and will not be fine tuned to the 
electroweak scale such that $H_2$ becomes sufficiently heavy to escape 
detection.

Below, we will first map out the parametric dependence of the shadow Higgs
mass on $g_s$, $\kappa$ and $r$. Then we will show that a light shadow Higgs is
viable from direct search and other experiments, and we give bounds on the
parameter space of the shadow Higgs from these experimental constraints. We 
will show that the shadow Higgs can be very light ($< 2m_e$), and we comment 
on the possibility of it been a dark matter candidate, and the degree of fine
tuning that is required.

\subsection{The mass of shadow Higgs}
To one-loop order, the mass of the heavy Higgs boson, $m_{H_2}$, is given by
the largest eigenvalue of the matrix defined in~\eqref{Eq:Msqto1L}. Due to its
complexity, we do not give its analytical form here. Rather, we will set
$m_{H_2}$ to the SM Higgs mass, and use it as a constraint which we solve
numerically for the allowed values of $\kappa$ to be used as an input to the
shadow Higgs mass, $m_{H_1}$.

\begin{figure}[htbp]
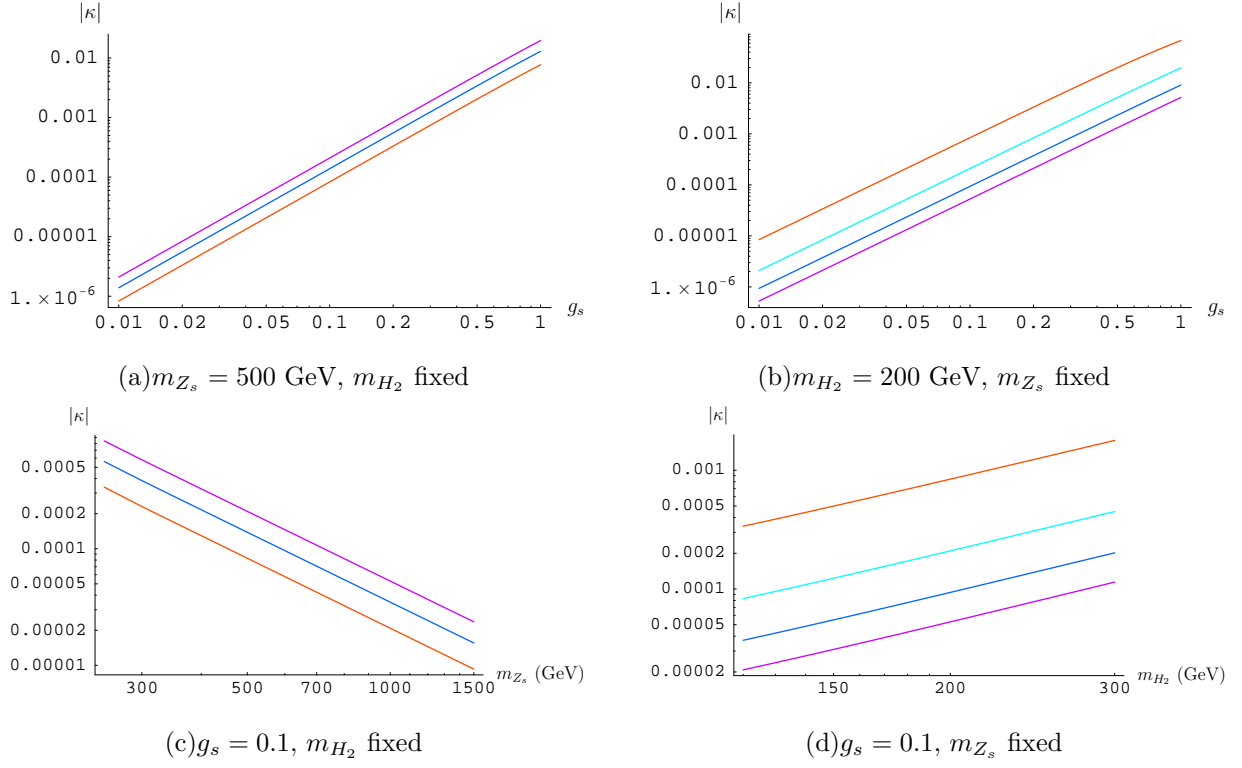

\centering
\subfigure[$m_{Z_s} = 500$~GeV, $m_{H_2}$ fixed]{
\label{subfig:kgs1}
\includegraphics[width=3in]{kgs_mZs500.epsi}}
\hspace{0.2in}
\subfigure[$m_{H_2} = 200$~GeV, $m_{Z_s}$ fixed]{
\label{subfig:kgs2}
\includegraphics[width=3in]{kgs_mH200.epsi}}
\subfigure[$g_s = 0.1$, $m_{H_2}$ fixed]{
\label{subfig:kmZs}
\includegraphics[width=3in]{kmZs_gsp1_log.epsi}}
\hspace{0.2in}
\subfigure[$g_s = 0.1$, $m_{Z_s}$ fixed]{
\label{subfig:kmH}
\includegraphics[width=3in]{kmH_gsp1_log.epsi}}
\caption{\label{Fig:gskmHmZs}
Contours of $|\kappa|$ as a function of $g_s$, $m_{H_2}$, and $m_{Z_s}$
individually.
In~\subref{subfig:kgs1} and~\subref{subfig:kmZs}, $m_{H_2}$ is fixed at
$120$~(bottom line), $160$, and $200$~(top line)~GeV.
In~\subref{subfig:kgs2} and~\subref{subfig:kmH}, $m_{Z_s}$ is fixed at
$250$~(top), $500$, $750$, and $1000$~(bottom)~GeV.}
\end{figure}

With $v_r$ fixed, $m_{H_2}$ depends on the parameters $g_s$, $\kappa$, and $r$
which we trade in for $m_{Z_s}$ using~\eqref{Eq:rse0}. In
Fig.~\ref{Fig:gskmHmZs}, we show the dependence of $\kappa$ on $g_s$,
$m_{Z_s}$, and $m_{H_2}$.
We see that $\kappa$ varies as a power of each of the parameters $g_s$, 
$m_{H_2}$, and $m_{Z_s}$. Note that $|\kappa|$ is at most of order $g_s^3$ in 
the range of $m_{Z_s}$ and $m_{H_2}$ we consider. In particular, the magnitude
of $\kappa$ decrease as the value of $m_{Z_s}$ ($m_{H_2}$) increase (decrease).

\begin{figure}[htbp]
\centering
\includegraphics[width=4in]{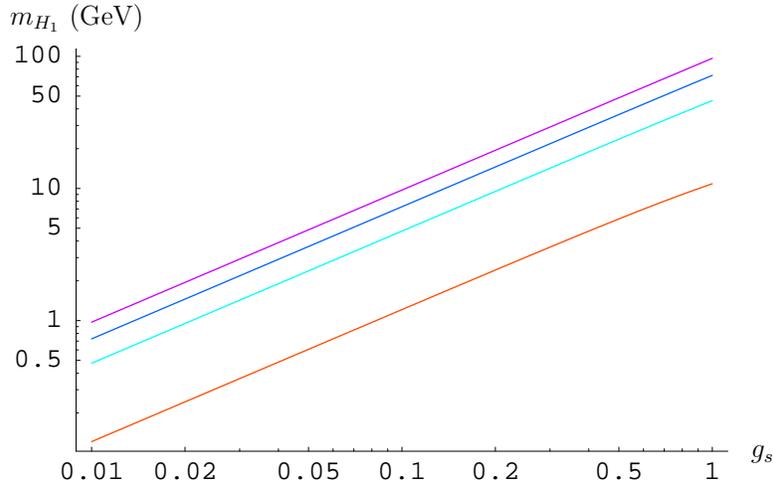}
\caption{\label{Fig:msmH}
The mass of shadow Higgs, $m_{H_1}$, as a function of $g_s$ for 
$m_{H_2} = 200$~GeV and $m_{Z_s}$ fixed at $250$~(bottom line), $500$, $750$, 
and $1000$~(top line)~GeV.}
\end{figure}

With $s_\eps$ taken to be zero, the shadow Higgs mass~\eqref{Eq:shmssq} takes 
the form
\begin{equation}\label{Eq:mssqse0}
m_{H_1}^2 = \frac{3v_r^2}{64\pi^2(1 + r)}\left[
\frac{3g_W^4}{2} + g_Y^2 g_W^2 + \frac{g_Y^4}{2} +
\frac{g_s^4 r^2}{2}\right] + \frac{v_r^2}{2\pi^2}(1 + r)\,\kappa^2
-\frac{3m_t^4}{2\pi^2 v_r^2(1 + r)} \,.
\end{equation}
In Fig.~\ref{Fig:msmH}, we show $m_{H_1}$ as a function of $g_s$ along the
contours $r(g_s)$ of constant $m_{Z_s}$ and $\kappa(g_s)$ of constant
$m_{H_2}$.
For $g_s \sim g_W \sim 0.65$, we see that the shadow Higgs mass varies from 
sub GeV range to tens of GeV, depending on $m_{Z_s}$. Note that $m_{H_1}$ 
decreases as a positive power of $g_s$, reflecting the fact that $\kappa^2$ 
decreases faster than $r^{-1} \propto g_s^2$ (see~\eqref{Eq:rse0}).

\subsection{Limits form direct LEP search}
The shadow Higgs couples to SM fields only through its mixing with the SM 
Higgs. From~\eqref{Eq:physH}, one can see that a triple coupling of the form
$H_1 F F$, where $F$ is a SM field, is simply that of the SM Higgs scaled by a
mixing factor of $n_1 = (1 + r)^{-1/2}$. At LEP, an important parameter used
in the direct Higgs search is $\xi^2 \equiv (g_{HZZ}/g_{HZZ}^{SM})^2$, where
$g_{HZZ}$ denotes the non-standard $HZZ$ coupling and $g_{HZZ}^{SM}$ that in
the SM. In terms of our model, the $\xi^2$ parameter becomes
\begin{equation}
\xi_1^2 = \left(\frac{g_{H_1ZZ}}{g_{HZZ}^{SM}}\right)^2
= \frac{1}{1 + r} \,, \qquad
\xi_2^2 = \left(\frac{g_{H_2ZZ}}{g_{HZZ}^{SM}}\right)^2
= \frac{r}{1 + r} \,,
\end{equation}
for the shadow Higgs, $H_1$, and the SM-like Higgs, $H_2$, respectively.

To see whether or not the shadow Higgs is ruled out at LEP, we can simply 
apply the LEP bound to $\xi_1^2$. The most stringent bound is obtained when 
the shadow Higgs mass is about 20~GeV, where
$\xi_1^2 \lesssim 2 \times 10^{-2}$~\cite{OPAL,LEP2}; elsewhere the bound is 
rather weak. From the discussion above, we have $g_{H_1ZZ} < g_{HZZ}^{SM}/10$ 
for much of the parameter space. Thus the shadow Higgs can easily pass the 
existing bound from the direct Higgs search.

\begin{figure}[htbp]
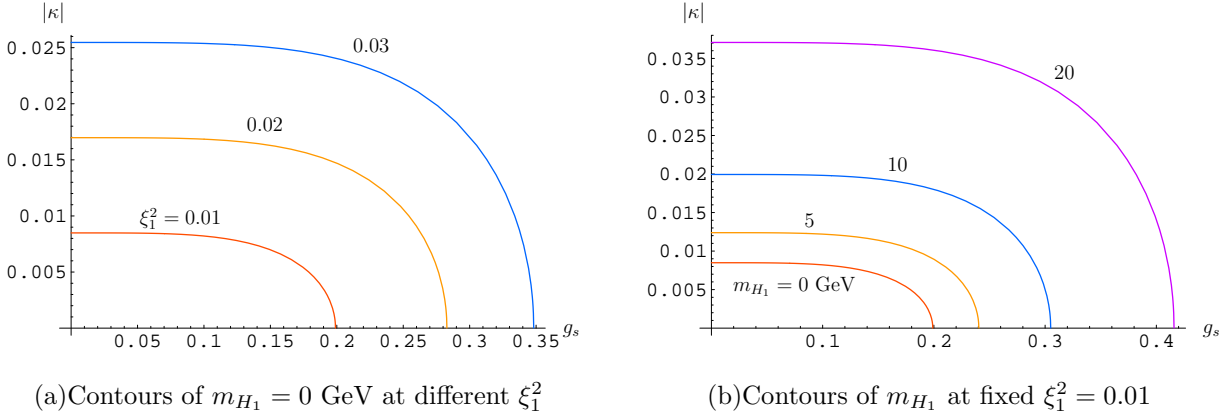

\centering
\subfigure[Contours of $m_{H_1} = 0$~GeV at different $\xi_1^2$]{
\label{subfig:ms0}
\includegraphics[width=3in]{ms0r.epsi}}
\hspace{0.2in}
\subfigure[Contours of $m_{H_1}$ at fixed $\xi_1^2 = 0.01$]{
\label{subfig:msrp2}
\includegraphics[width=3in]{msrp2.epsi}}
\caption{\label{Fig:msC}
Contours of $m_{H_1}$ as a function of $g_s$ and $\kappa$ for fixed values
of $\xi_1^2 = (1+r)^{-1}$.}
\end{figure}

\begin{figure}[htbp]
\centering
\includegraphics[width=4in]{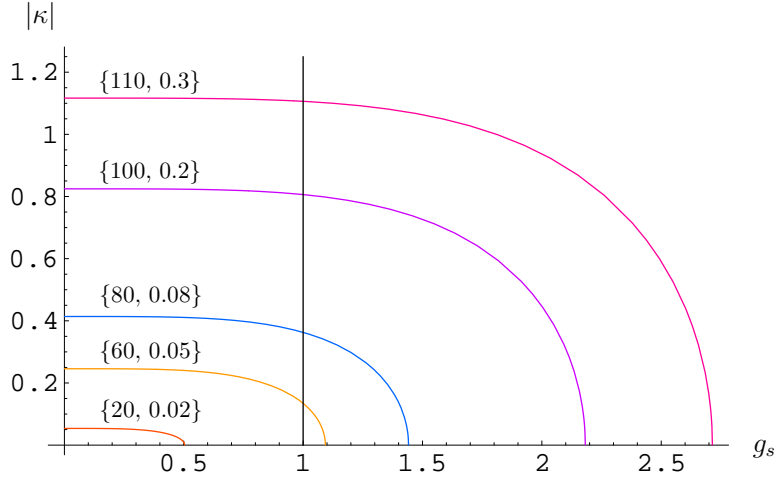}
\caption{\label{Fig:msbound}
Contours of $m_{H_1}$ at various $\{m_{H_1}\,(\mathrm{GeV})\,,\xi_1^2\}$ for 
$\xi_1^2$ the maximum value of the LEP $95\%$ CL upper bound found in 
Ref.~\cite{LEP2}. The vertical line marks the boundary for $g_s < 1$.}
\end{figure}

With a light shadow Higgs ($m_{H_1} < m_H^{SM}$) viable, we show in 
Fig.~\ref{Fig:msC} and Fig.~\ref{Fig:msbound} the parameter space that can be 
constrained by the mass of the shadow Higgs, $m_{H_1}$, and the ratio of the 
scalar-$Z$ triple coupling squared, $\xi_1^2$. Fig.~\ref{subfig:ms0} shows 
that for fixed $m_{H_1}$, decreasing $\xi_1^2$ (or equivalently, increasing 
$r$) shrinks the contour of constant shadow Higgs mass, while 
Fig.~\ref{subfig:msrp2} shows that for fixed $\xi_1$ ($r$), increasing the 
value of $m_{H_1}$ expands the contour. Given these, the contours in 
Fig.~\ref{Fig:msbound} forms the upper bound on the allowed values of 
$\{g_s,\kappa\}$ for each fixed value of $m_{H_1}$.

\subsection{Other limits}
The one-loop contribution to muon g-2 from the (neutral) SM Higgs is well
known~\cite{higgshunter}. From~\eqref{Eq:physH}, scaling it by a factor of
$(1+r)^{-1}$ gives the contribution due to the shadow Higgs
\begin{equation}
\tri a_\mu = \frac{1}{1+r}\,\frac{G_F\,m_\mu^2}{4\pi^2\sqrt{2}}\,
I\!\left(\frac{m_{H_1}^2}{m_\mu^2}\right) \,,
\end{equation}
where
\begin{equation}
I(x)= \int^1_0\!\!dy\,\frac{y^2(2-y)}{(1-y)x + y^2} \sim
\begin{cases}
\frac{3}{2}-\pi\sqrt{x} \,, & x \ll 1 \\
\frac{1}{x}(\log{x}-\frac{7}{6}) \,, & x \gg 1 \\
0 \,, & x \rightarrow \infty
\end{cases} \,.
\end{equation}
Thus, for $m_{H_1} > 1$~GeV,
\beq
\tri a_\mu \sim \frac{2.5\times 10^{-11}}{1+r}
\left(\frac{\mathrm{GeV}}{m_{H_1}}\right)^2 \,,
\eeq
while for $m_{H_1} \ll m_\mu$, $\tri a_\mu \sim 3/(1+r)\times 10^{-9}$. Now the
latest muon g-2 measurement gives
$\tri a_\mu^{exp}-\tri a_\mu^{SM} =2.8(8)\times 10^{-9}$~\cite{Hertzog:2006sc},
we see that the shadow Higgs contribution to muon g-2 is at most $(1+r)^{-1}$ 
of that difference. Giving that $r>10$ in most of the parameter space, muon g-2
gives no constraint on the shadow Higgs mass.

For a light shadow Higgs with $m_{H_1} \leq 1$~GeV, the most stringent 
constraint comes from the $B$ meson decays. From comparing the $b \ra s\,H_1$ 
penguin diagram to the tree-level $b \ra c\,W$ transition, one gets an 
inclusive branching ratio relation \cite{higgshunter}
\beq
\frac{\Gamma(B\ra H_1 X)}{\Gamma(B\ra e\nu X)} \sim
\frac{2.95}{1+r}\left(\frac{m_t}{M_W}\right)^4
\left(1-\frac{m_{H_1}^2}{m_b^2}\right)^2
\left|\frac{V^*_{ts}V_{tb}}{V_{cb}}\right|^2 \,,
\eeq
where the numerical factor contains the phase space difference of the final
state $c$ and $s$ quarks. Taking $Br(B \ra e\nu X)=0.123$ and 
$m_{H_1} \ll m_b$, we get
\beq
Br(B \ra H_1 X) \sim \frac{8}{1+r} \,.
\eeq

In order to make comparisons with the experimental bound on the exclusive decay
modes of the $B$ meson, the shadow Higgs decay branching ratios are needed. 
However, since the shadow Higgs can decay into light hadrons, the branching 
ratio calculations involve many hadronic uncertainties. Consider, for example,
a shadow Higgs with $m_{H_1} = 500$~MeV that decays mainly into two pions and 
$\mu^+\mu^-$. With the help of chiral perturbation theory, the branching ratio
$Br(H_1 \ra \mu^+\mu^-)$ is estimated to be $\sim 30\%$~\cite{OCRMW06}.
Now if the shadow Higgs is heavier than $2m_K$ or $2m_\tau$, this and the whole
decay pattern will change dramatically. Moreover, chiral perturbation may
not be reliable anymore in these cases to calculate the decay widths.

From Ref.~\cite{PDG}, $Br(B \ra \mu^+\mu^- X) < 3.2 \times 10^{-4}$. Suppose
the shadow Higgs decays only into $\mu^+\mu^-$, then a (rather) conservative 
lower bound on $r$ is given by
\beq
2 \times 10^4 < r\left(1 - \frac{m_{H_1}^2}{m_b^2}\right)^{-2} \,.
\eeq
Given this bound, the quarkonium decays branching ratios
$Br(J/\Psi \ra H_1\gamma) < 10^{-9}$ and
$Br(\Upsilon \ra H_1\gamma) = 1.8 \times 10^{-4}/(1+r) \leq 10^{-8}$,
which involve tree-level processes, become insignificant in comparison with
$Br(B\ra \mu^+\mu^- X)$, which involves a one-loop process.

\subsection{The case of an extremely light shadow Higgs ($m_{H_1} < 2m_e$)}
When the shadow Higgs is lighter than $2 m_e$, it decays almost completely 
into two photons. The corresponding effective interaction can be derived by 
summing up the contributions from having the $t$-quark and the W-boson running
in the loop, and is given by~\cite{higgshunter}
\begin{equation}\label{Eq:s2photon}
\tri {\cal L} = \frac{g_{H_1\gamma\gamma}}{4} F^{\mu\nu}F_{\mu\nu} H_1 \,,
\qquad
g_{H_1\gamma\gamma} = \frac{7\alpha}{3\pi v_r(1+r)} \sim
\frac{2.2 \times 10^{-5}(\mathrm{GeV})^{-1}}{1+r} \,,
\end{equation}
where $\alpha$ is the fine-structure constant.

There is currently no direct experimental bound on the coupling constant
$g_{H_1\gamma\gamma}$. The shadow Higgs and one of the two photons in the 
effective operator can be attached to charged fermions which yield a one-loop
contribution to the magnetic moment. However this contribution is buried deep
inside the one-loop g-2 contribution discussed above.

The lifetime of the shadow Higgs can be estimated to be
\begin{equation}\label{Eq:gsgg}
\tau_{H_1} \sim (1+r)\left(\frac{68\,\mathrm{keV}}{m_{H_1}}\right)^3 sec 
\sim (1+r)\left(\frac{0.1\,\mathrm{eV}}{m_{H_1}}\right)^3 10^{10}\,yr \,.
\end{equation}
For a shadow Higgs lighter than a few tens of keV, its life time may be long
enough for it to escape and carry away energy from stars in the horizontal 
branch with a typical radius of a few tens of light second. Recently, an upper
bound on the coupling of a very light exotic spin-0 particle to two photons
has been placed at $1.1 \times 10^{-10}\,\mathrm{GeV}^{-1}$ by the CAST
Collaboration~\cite{CAST}. Applying this bound to the scalar 
case~\eqref{Eq:gsgg} implies that $r > 10^5$.

For an even lighter shadow Higgs, the stellar energy lost through 
$e\,\gamma \ra e\,H_1$ puts a stronger bound on the electron-shadow Higgs 
Yukawa coupling, $(y^2_{e H_1})/4\pi\leq 10^{-29}$~\cite{Raffelt}. But this 
would push our model into an extremely fine-tuned region, where 
$r \geq 10^{16}$.

Recall that $r > 10^4$ is already necessary when considering the rare
$B \ra \mu^+\mu^- X$ decay. In this region, if the shadow Higgs is lighter than
$0.1\,r^{1/3}$~eV, which is $2.15$~eV for $r=10^4$ and $215$~eV for 
$r=10^{10}$, it can have a cosmologically interesting life time and may
contribute a noticeable fraction to the dark matter density. Note that this
does not require one to impose a discrete symmetry such as an extra $Z_2$
parity as in Ref.~\cite{BF04}.

\subsection{Searching for the shadow Higgs at the LHC}
\begin{figure}[htbp]
\centering
\includegraphics[width=2in]{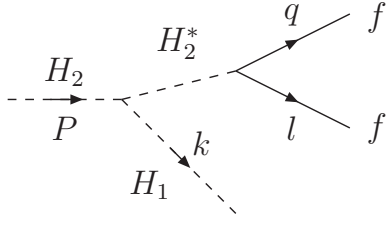}
\caption{\label{Fig:3Bdecay}
Leading order 3-body $H_2 \ra H_1 f\bar{f}$ decay.}
\end{figure}

As can be seen from~\eqref{Eq:physH}, the 2-body decay widths of the SM-like
Higgs $H_2$ are simply that of the SM scaled by a factor $n_2^2 = r/(1+r)$,
and so no significant changes are expected here. More interesting are the
3-body decays described in Fig.~\ref{Fig:3Bdecay},
whose amplitude is given by
\beq
\mathcal{M} = \frac{y_f\,\lambda_3\,v_r}{(P-k)^2-m_{H_2}^2}\,\bar{u}(l)\,v(q)
\,, \qquad
\lambda_3 \equiv 4\kappa\sqrt{1+r} \,,
\eeq
where $y_f$ is the SM Higgs-fermion Yukawa coupling, and $-\lambda_3 v_r/2$ is
the coupling of the $H_2 H_2 H_1$ vertex. From a standard calculation, and the 
fact that the $b$-quark is much lighter than the SM Higgs, the decay width 
reads
\begin{equation}\label{Eq:H2Width}
\Gamma(H_2 \ra H_1 f\bar{f}) =
N_c\frac{y_f^2\,\lambda_3^2}{128\pi^3}\frac{v_r^2}{m_{H_2}}F_H(\beta) \,,
\end{equation}
where
\begin{align}\label{Eq:FH}
F_H(\beta) &= \int^{1+\beta^2}_{2\beta}\!\!\!dx\,
\frac{1+\beta^2-x}{(x-\beta^2)^2}\sqrt{x^2-4\beta^2} \,, \qquad
x = \frac{2k_0}{m_{H_2}} \,, \quad \beta = \frac{m_{H_1}}{m_{H_2}} \,,
\notag \\
&\sim -2-\log\beta +\frac{5\pi}{4}\beta + \mathcal{O}(\beta^2) \,,
\qquad 0 < \beta \ll 1 \,.
\end{align}
We have thus
\begin{equation}\label{Eq:H2Incl}
\sum_f \Gamma(H_2 \ra H_1 f \bar{f}) =
\frac{\lambda_3^2}{64\pi^3 m_{H_2}}F_H(\beta)\sum_f N_c m_f^2 \sim
0.3\lambda_3^2\,F_H(\beta)\left(\frac{120\mathrm{GeV}}{m_{H_2}}\right)
\mathrm{MeV} \,,
\end{equation}
where the sum runs over $b$, $c$ and $\tau$.

The inclusive width given in~\eqref{Eq:H2Incl} is to be compared with
$\Gamma_{total} \sim 40$~MeV for a SM Higgs of $120$~GeV.
Table~\ref{tab:H2Gamma} lists the relevant quantities entering
into~\eqref{Eq:H2Incl} for $m_{H_2} = 120$~GeV, $m_{Z_s} = 500$~GeV and
$m_{H_1} = 0.001,\,1,\,30$~GeV. We see that the tree-level $H_2 H_2 H_1$ 
coupling (in units of $v_r$) is tiny in all cases. Thus, we expect the 3-body 
decay process, $H_2 \ra H_1 f\bar{f}$, to have little impact on the branching 
ratios of the SM-like $H_2$.
\begin{table}[hbp]
\caption{\label{tab:H2Gamma} Values of $F_H$ and $\lambda_3$ at various
$m_{H_1}$ for $m_{H_2} = 120$~GeV and $m_{Z_s} = 500$~GeV.}
\begin{tabular}{|c|c|c|c|}
\hline
$m_{H_1}$~(GeV) & $0.001$ & $1$ & $30$ \\
\hline
$F_H$ & $9.96$ & $2.82$ & $0.168$ \\
\hline
$\lambda_3$ & $-1.4 \times 10^{-6}$ & $-1.4 \times 10^{-3}$ & $-0.04$ \\
\hline
\end{tabular}
\end{table}

\begin{figure}[htbp]
\centering
\includegraphics[width=4in]{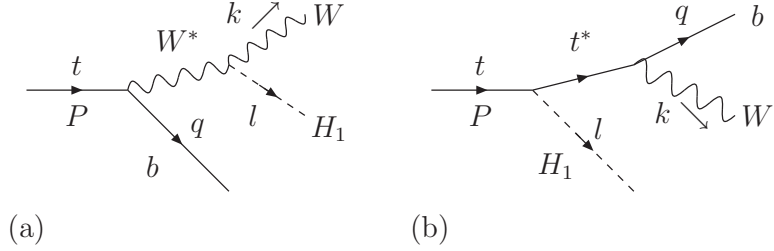}
\caption{\label{Fig:top3Bdecay}
Leading order 3-body $t \ra H_1\,b\,W^+$ decay.}
\end{figure}

Since the Yukawa coupling of the top to the SM Higgs is the largest amongst the
fermions, it would also be the largest fermion-shadow Higgs coupling as well. 
Thus we expect that there is a good chance of detecting the shadow Higgs in 
precision top decay studies such as at the LHC, where $8 \times 10^6\,t\bar{t}$
events per year are expected at a luminosity of 
$10^{33}\,cm^{-1}s^{-1}$~\cite{ATLAS}.

As is the case with the SM Higgs search at the LHC, the primary process that
can reveal the presence of the shadow Higgs is the 3-body decay, 
$t \ra H_1\,b\,W^+$, described in Fig.~\ref{Fig:top3Bdecay}. Taking 
$m_t=174$~GeV, $m_W=80.4$~GeV, $m_b=4.5$~GeV, and $y_t \sim 1.0$, after a 
standard but tedious calculation the decay width is evaluated to be
\begin{equation}\label{Eq:tWidth}
\Gamma(t \ra H_1\,b\,W^+) \sim 
\{16.85,\,4.78,\,0.221\}\times\frac{\sqrt{2}G_F}{256\pi^3}\frac{m_t^3}{1+r} 
\sim \{18,\,5,\,0.2\}\times\frac{ 10^{-2}}{1+r}\,\mathrm{GeV} \,,
\end{equation} 
for $m_{H_1} = 0.001,\,1,\,30$~GeV respectively. This is to be compared with 
the top decay width predicted in the SM~\cite{Jezabek:1988iv}
\beq
\Gamma_t = \frac{G_F M_t^3}{8\pi\sqrt{2}}(1-\eta^2)^2(1 + 2\eta^2)
\left[1-\frac{2\alpha_s}{3\pi}\left(\frac{2\pi^2}{3}-\frac52\right)\right]
= 1.37\,\mathrm{GeV} \,, 
\eeq
where $\eta = m_W/m_t$ and $\alpha_s(m_Z) = 0.118$. 

Suppose that $m_{H_1}=30$~GeV, $r \simeq 10$, and the experimental sensitivity
can reach down to $10^{-4}$ (which is expected from the LHC at high 
luminosity), then~\eqref{Eq:tWidth} suggests that the presence of the shadow 
Higgs can be tested by studying the top decay width. However, as discussed 
before, the parameter space for having $r \simeq 10$ is small. From 
Fig.~\ref{subfig:rmZsgs1}, we see that for $m_{Z_s} > 500$~GeV, $g_s > 1$ is 
required, and so the use of perturbation theory becomes questionable. Only for
$m_{Z_s} < 300$~GeV can a perturbative $g_s$ be easily maintained. Since $r$ 
is more naturally of $\mathcal{O}(100)$, a search for the shadow Higgs may 
require the LHC to operate at high luminosity for extended periods of time.

\section{Summary}\label{Sec:Summ}
We have studied the scale invariant version of a hidden extra $U(1)$ model
with radiative gauge symmetry breaking. The dimensional transmutation
mechanism results in a heavy scalar which we identify as the SM Higgs ($H_2$),
and a light scalon which we call the shadow Higgs ($H_1$). There are no
other physical spin-0 particle in the model.

Unlike other extended Higgs models, there is no tree level $H_2 H_1 H_1$
couplings. Thus, the model predicts no additional two body decays for $H_2$,
and to leading order, the SM Higgs physics is only modified by a factor of
$r/(1+r)$. As for the shadow Higgs, it behaves in general like a lighter
version of the SM Higgs with couplings to quarks and gauge bosons reduced by a
factor of $1/(1+r)$. Phenomenological considerations from LEP constraints
dictate that $r>10$ for $m_{H_1} < 100$~GeV.

For a shadow Higgs with mass in the range $2m_e < m_{H_1} < 1$~GeV, the most
stringent constrain comes from the $B\rightarrow \mu^+\mu^- X$ decay that
leads to a lower limit of $r > 10^4$. For $m_{H_1} < 20$~keV, stellar cooling
imposes the limit of $r > 10^5$. While not impossible, this stretches the
limit of fine tuning. For a cosmologically interesting shadow Higgs,
$r > 10^6$ would be required.

Given that the coupling of the shadow Higgs to SM particles will be quite weak,
the shadow Higgs will be elusive to most searches. However, if its mass is in
the range of a few to 100~GeV, it can be detected in top decays. In particular,
there will a parallel mode alongside the $t \rightarrow H_2\,W\,b$ decay in
which the SM-like Higgs is replaced by the lighter shadow Higgs. If
$r \simeq 10$, we expect a branching ratio of $\mathcal{O}(10^{-4})$ for the 
shadow Higgs, which should be detectable at the LHC with high luminosity runs. 
In the event that the SM-like Higgs is heavier than or is too close to $m_t$ 
so that the decay is kinematically suppressed, the shadow Higgs will be the 
only such decay to be seen. This search can be extended to the ILC where the 
environment will be much cleaner.

\section{acknowledgements}
The research of J.N.N. and J.M.S.W. are partially supported by the Natural
Science and Engineering Council of Canada. The research of W.F.C. is supported
by Taiwan NSC under grant 95-2112-M-007-032.


\begin{thebibliography}{99}
\bibitem{SW05}
R.~Schabinger and J.~D.~Wells,
Phys.\ Rev.\ D {\bf 72}, 093007 (2005).

\bibitem{CFW05}
S.~Chang, P.~J.~Fox and N.~Weiner,
JHEP {\bf 0608}, 068 (2006).

\bibitem{SZ06a}
M.~J.~Strassler and K.~M.~Zurek,
arXiv:hep-ph/0604261.

\bibitem{PW06}
B.~Patt and F.~Wilczek,
arXiv:hep-ph/0605188.

\bibitem{MW06}
A.~V.~Manohar and M.~B.~Wise,
Phys.\ Rev.\ D {\bf 74}, 035009 (2006).

\bibitem{CDU06}
D.~G.~Cerdeno, A.~Dedes and T.~E.~J.~Underwood,
JHEP {\bf 0609}, 067 (2006).

\bibitem{OCRMW06}
D.~O'Connell, M.~J.~Ramsey-Musolf and M.~B.~Wise,
arXiv:hep-ph/0611014.

\bibitem{BTGR06}
O.~Bahat-Treidel, Y.~Grossman and Y.~Rozen,
arXiv:hep-ph/0611162.

\bibitem{MN06}
K.~A.~Meissner and H.~Nicolai,
arXiv:hep-th/0612165.

\bibitem{EQ07}
J.~R.~Espinosa and M.~Quiros,
arXiv:hep-ph/0701145.

\bibitem{ExtraZ}
see, e.g.
R.~W.~Robinett and J.~L.~Rosner,
Phys.\ Rev.\ D {\bf 25}, 3036 (1982),
D {\bf 27}, 679 (1983) {\bf (E)};
P.~Langacker, R.~W.~Robinett and J.~L.~Rosner,
Phys.\ Rev.\ D {\bf 30}, 1470 (1984);
F.~Zwirner,
Int.\ J.\ Mod.\ Phys.\ A {\bf 3}, 49 (1988).

\bibitem{EL99}
J.~Erler and P.~Langacker,
Phys.\ Lett.\ B {\bf 456}, 68 (1999).

\bibitem{ADH02}
T.~Appelquist, B.~A.~Dobrescu and A.~R.~Hopper,
Phys.\ Rev.\ D {\bf 68}, 035012 (2003).

\bibitem{KW06}
J.~Kumar and J.~D.~Wells,
Phys.\ Rev.\ D {\bf 74}, 115017 (2006).

\bibitem{CNW06}
W.~F.~Chang, J.~N.~Ng and J.~M.~S.~Wu,
Phys.\ Rev.\ D {\bf 74}, 095005 (2006).

\bibitem{BHR06}
R.~Barbieri, L.~J.~Hall and V.~S.~Rychkov,
Phys.\ Rev.\ D {\bf 74}, 015007 (2006).

\bibitem{BF04}
V.~Silveira and A.~Zee,
Phys.\ Lett.\ B {\bf 161}, 136 (1985);
C.~Boehm and P.~Fayet,
Nucl.\ Phys.\ B {\bf 683}, 219 (2004).

\bibitem{CW73}
S.~R.~Coleman and E.~Weinberg,
Phys.\ Rev.\ D {\bf 7}, 1888 (1973).

\bibitem{Hempf96}
R.~Hempfling,
Phys.\ Lett.\ B {\bf 379}, 153 (1996).

\bibitem{LEP2}
R.~Barate {\it et al.}  [LEP Working Group for Higgs boson searches],
Phys.\ Lett.\ B {\bf 565}, 61 (2003).

\bibitem{GW76}
E.~Gildener and S.~Weinberg,
Phys.\ Rev.\ D {\bf 13}, 3333 (1976).

\bibitem{Future}
In preparation.

\bibitem{OPAL}
P.~D.~Acton {\it et al.}  [OPAL Collaboration],
Phys.\ Lett.\ B {\bf 268}, 122 (1991).

\bibitem{higgshunter}
J.~F.~Gunion, H.~E.~Haber, G.~L.~Kane and S.~Dawson,
{\it The Higgs Hunter's Guide},
Perseus Books Group, 2000.

\bibitem{Hertzog:2006sc}
D.~W.~Hertzog,
arXiv:hep-ex/0611025.

\bibitem{PDG}
W.~M.~Yao {\it et al.}  [Particle Data Group],
J.\ Phys.\ G {\bf 33}, 1 (2006).

\bibitem{CAST}
K.~Zioutas {\it et al.}  [CAST Collaboration],
Phys.\ Rev.\ Lett.\  {\bf 94}, 121301 (2005).

\bibitem{Raffelt}
G.~G.~Raffelt,
{\it Stars As Laboratories For Fundamental Physics: The Astrophysics Of Neutrinos, Axions, And Other Weakly
Interacting Particles},
Chicago Univ. Pr., Chicago, 1996.

\bibitem{ATLAS}
ATLAS: Detector and physics performance technical design report,
Volume 1,
CERN-LHCC-99-14.

\bibitem{Jezabek:1988iv}
M.~Jezabek and J.~H.~Kuhn,
Nucl.\ Phys.\ B {\bf 314}, 1 (1989).
\end{thebibliography}
\end{document}